# The Equilibrium Dynamics of Thermostatic End-Use Load Diversity as a Function of Demand

D. P. Chassin, *Member IEEE* & J. M. Malard

*Abstract* – **A recently developed state-of-the-art agent-based simulation of power distribution systems is capable of modeling the thermal and demand response behavior of many thousands of thermostatic end-use loads. It computes the total power consumed by electric power customers when responding to real-time prices. Investigation of the initialization transients has led to a more precise understanding of the relationship between the end-use state probability of thermostatic loads, end-use demand, and load diversity.**

## I. Introduction

Advanced market-based load control provides numerous economic benefits to utilities, electric generation and transmission operators, and their customers [1]. Furthermore, load control has long been considered an effective strategy in mitigate price volatility and its adverse effects on electricity markets [2]. However, the interactions between markets and loads are not well understood phenomena. Traditionally three major types of price responsive load control strategies have been discussed: curtailment [3], substitution [4], and storage [5]. But more recently load shifting end-uses that cycle naturally has been considered more carefully because a) it is less likely to adversely impact the customer, particularly when done over relatively short time frames [11], and b) it is more readily implemented as a true distributed control strategy that require no centralized dispatch [6]. The key advantage of load shifting is that it contributes positive to overall load leveling with system-wide economic benefits [1].

The Power Distribution System Simulator (PDSS) was developed at Pacific Northwest National Laboratory to study the benefits and challenges of this type of advanced, market-based load control strategies [7]. PDSS is a state-of-the-art high-performance agent-based simulation capable of individually modeling many thousands of end-use loads with a wide variety control strategies, thermal models, and demand models. Demand behavior models were developed from ELCAP load-shapes [8] and the thermal models use the Equivalent Thermal Parameters (ETP) model [9].

The computational resources the PDSS require for large population studies are quite substantial and it has been an objective of the development to minimize the simulation duration by a variety of strategies including reduced initialization processing, use of high-performance computing clusters [10], and increased use of advanced reduced-order models [11]. The focus of this paper is on the development of a reduced-order model of load diversity that is more accurate and better able to responds to price signals than the standard models of load diversity.

## II. Classical Model of State Diversity

PDSS makes the assumption that under steady-state conditions the probability density function for the temperature and mode of thermostatic end-use loads is based solely on the heating and cooling rates of the stored thermal mass [7]. However, three parameters are known to govern the behavior of thermostatic loads: the duty cycle, the cycling period, and the end-use demand [11].

The duty cycle $\varphi$ is estimated based on the Duty Cycle Model [12], which states that

$$\varphi = t_{on} / t \tag{1}$$

where $t_{on}$ is the on time during one cycle, and $t$ is the total cycling time $t = t_{on} + t_{off}$. If $P_{avg}$ is the average connected load and $P_{max}$ is the total connected load, then we can estimate the average connected load $P_{avg}$ based on the fraction of on time $t_{on}$ with respect to the total time $t$, such that $P_{avg} = P_{max} \times t_{on} / t$. We find that

$$\varphi = \frac{P_{avg}}{P_{\max}} \tag{2}$$

The duty cycle is a dimensionless number between zero and one. In cases where $t_{on}$ and $t_{off}$ are not known *a priori*, they can be estimated from the heating and cooling rates $r_{on}$ and $r_{off}$ [1], such that

$$r_{on} = \frac{T_{off} - T_{on}}{t_{on}} \quad , \quad r_{off} = \frac{T_{on} - T_{off}}{t_{off}} \tag{3}$$

where $T_{on}$ and $T_{off}$ are the thermostat's true on and off temperatures, which includes the dead-band around the set-point. These values can be computed from the device's thermal properties, e.g., heat capacity, heat loss or gain coefficients.

The period $t$ for the cycle of an arbitrary thermostatic appliance is given by

$$t = t_{on} + t_{off} = \left(T_{off} - T_{on}\right)\left(\frac{1}{r_{on}} - \frac{1}{r_{off}}\right) \tag{4}$$

The signs of $r_{on}$ and $r_{off}$ must be consistent with the respective positions of $T_{on}$ and $T_{off}$, i.e., heating appliances have $T_{on} < T_{off}$, $r_{on} > 0$, and $r_{off} < 0$ and cooling appliances have $T_{on} > T_{off}$, $r_{on} < 0$, and $r_{off} > 0$. The fundamental period of a thermostatic load is dependent only on the thermal and control properties of the load. Thus it would be more accurate to assert that the distribution of states depends on the thermal properties, the control parameters, and the demand.

Some authors prefer to use diversity factors $k_d$ when describing the aggregate effect of randomly distributed individual loads on distribution circuit loading calculations [13]. *IEEE Standard 141-1993* defines diversity factor in §2.4.1.3.5 as follows:

> «*The ratio of the sum of the individual non-coincident maximum demands of various subdivisions of the system to the maximum demand of the complete systems. The (unofficial) term* diversity*, as distinguished from* diversity factor *refers to the percent of time available that a machine [...] has its maximum or nominal load or demand [...].*»

We note that *diversity* is defined here as the duty cycle discussed above. It has been shown that for thermostatic devices, the diversity factor is computed as $k_d = t \,/\, t_{on}$, which we observe is the inverse of the duty cycle [11]. As a matter of convention we refer to duty cycle when discussing the property of a single load or a class of similar loads, and we refer to diversity and diversity factor when discussion the aggregate behavior of populations of loads.

## III. Discrepancies of Classical Model

Thermostatic loads are readily modeled using a state space in which two dimensions are considered: temperature and mode [7] [11] [14]. Indeed, the first version of PDSS implements such an agent-based model of many thousands of loads, operating independently in response to real-time price signals posted at arbitrary times, which are in turn updated in response to the load. In order to eliminate the simulation's initial settling time, the initial distribution of the thermostatic devices over the state space must be the steady-state condition. It was therefore desirable to compute the probability density function for the steady-state condition for an arbitrary thermostatic device. The lack of discussion of steady-state distributions in the load modeling literature suggested to the designers of PDSS that the distribution was generally regarded as obvious and intuitive: if the demand and period were constant, then the duty cycle was the only governing factor and both the steady-state and initial probability density functions for thermostatic devices must be uniform, such that

$$\Pr\{x \le T < x + dx\} = \begin{cases} x < T_{on} & : \quad 0 \\ T_{on} \le x \le T_{off} & : \quad \dfrac{dx}{\left|T_{off} - T_{on}\right|} \\ x > T_{off} & : \quad 0 \end{cases} \tag{5}$$

$$\Pr\{on\} = \varphi \quad , \quad \Pr\{off\} = 1 - \varphi$$

However, in the course of testing the first version of PDSS, two problems were observed whenever the demand was initially non-zero, as illustrated in Fig. 1. These problems called into question the load diversity model employed. As a result, an investigation was conducted to derive corrections that could eliminate the initial transient and erroneous diversity prediction. The results of this investigation have provides new insight into the behavior of thermostatic loads in general.

[1] In the case of cooling appliances, the heating rate is $r_{off}$ and the cooling rate $r_{on}$.

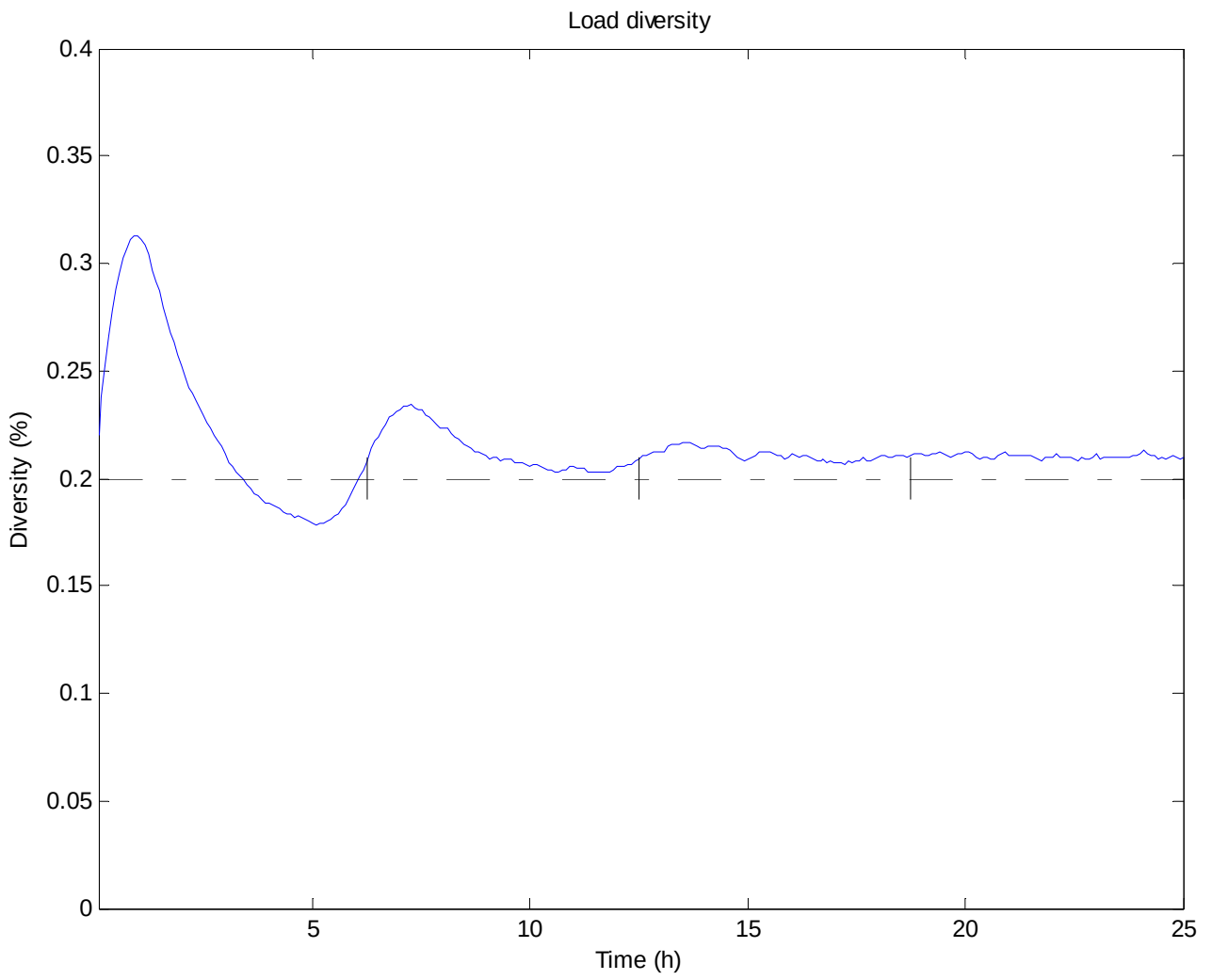

Fig. 1: The load diversity at equilibrium is not as predicted by the classical model (dash-dot) when the demand is non-zero. The tick marks indicated the predicted period. Note also the initialization transient.

A simplified version of the PDSS model was constructed in MATLAB, tested and examined to determine what the true steady-state distribution might be. The initial distribution used by PDSS was provided to the verification simulation and is shown in Fig. 1. It should also be noted that this model is a continuous model, much like that which the PDSS agents use, and not like the discrete model used in the consideration of the diversity impact of price-responsive thermostatic end-uses [11].

The result of the simulations shows that the final steady-state distributions differ depending on whether the demand is zero. When the demand is zero, the steady-state distribution remains uniform, just as it was at initial conditions, as shown in Fig. 2. However, Fig. 3 reveals a phenomenon that has not been previously described: the steady-state distribution is not uniform when demand is non-zero, rather it follows what appears to be an exponential curve. This result suggests that the original assumption that demand usage did not greatly affect the steady-state distribution was incorrect.

In addition, according to the standard model the diversity of studied scenario was expected to be 0.2. However, when the demand is non-zero, the diversity exceeds the predicted value by approximately 10%, suggesting that the load diversity factor is closer to unity than one might expect based solely on knowledge of the thermal properties of the end-uses.

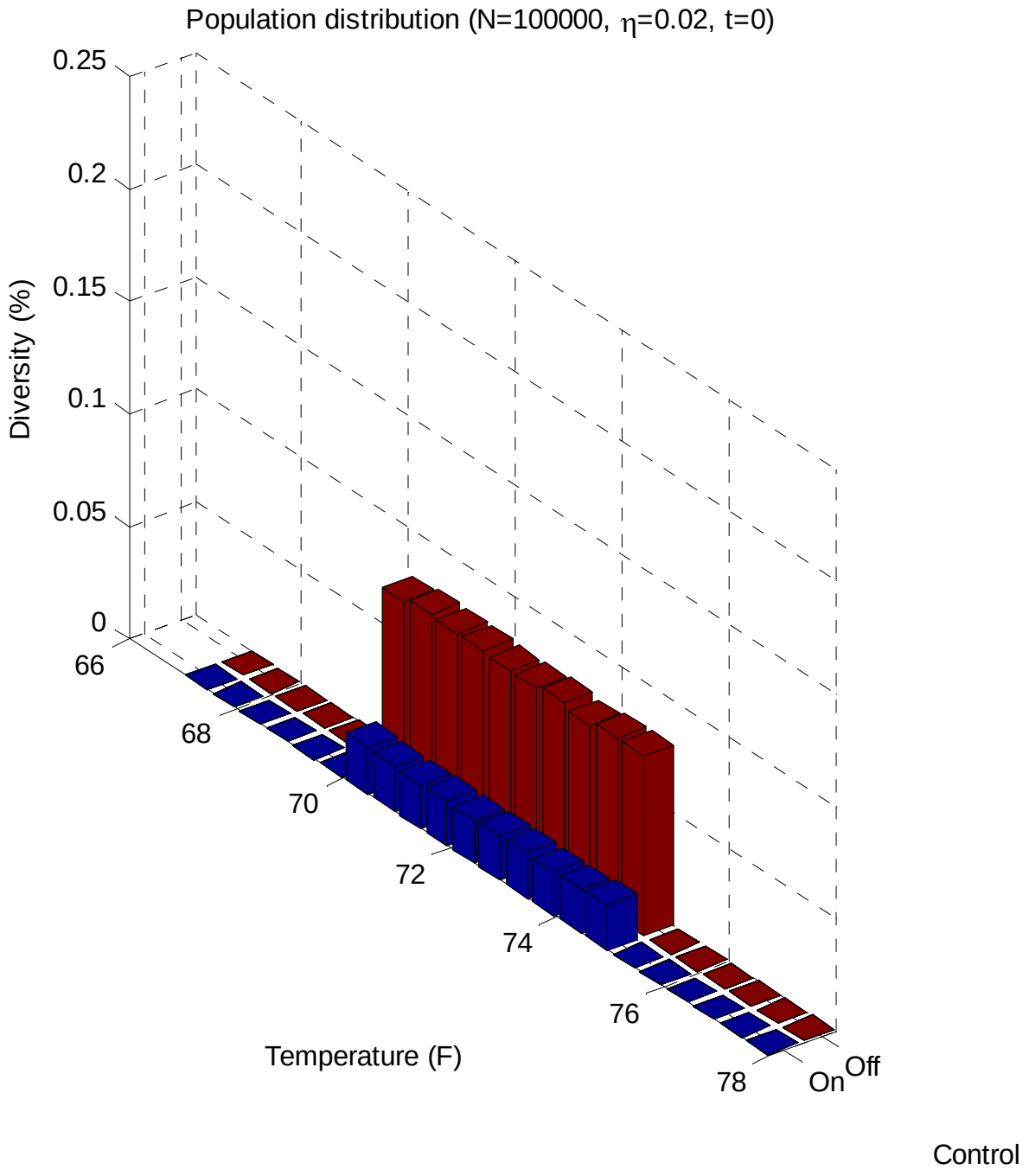

Fig. 2: The initial state distribution is uniform and remains uniform when demand is zero.

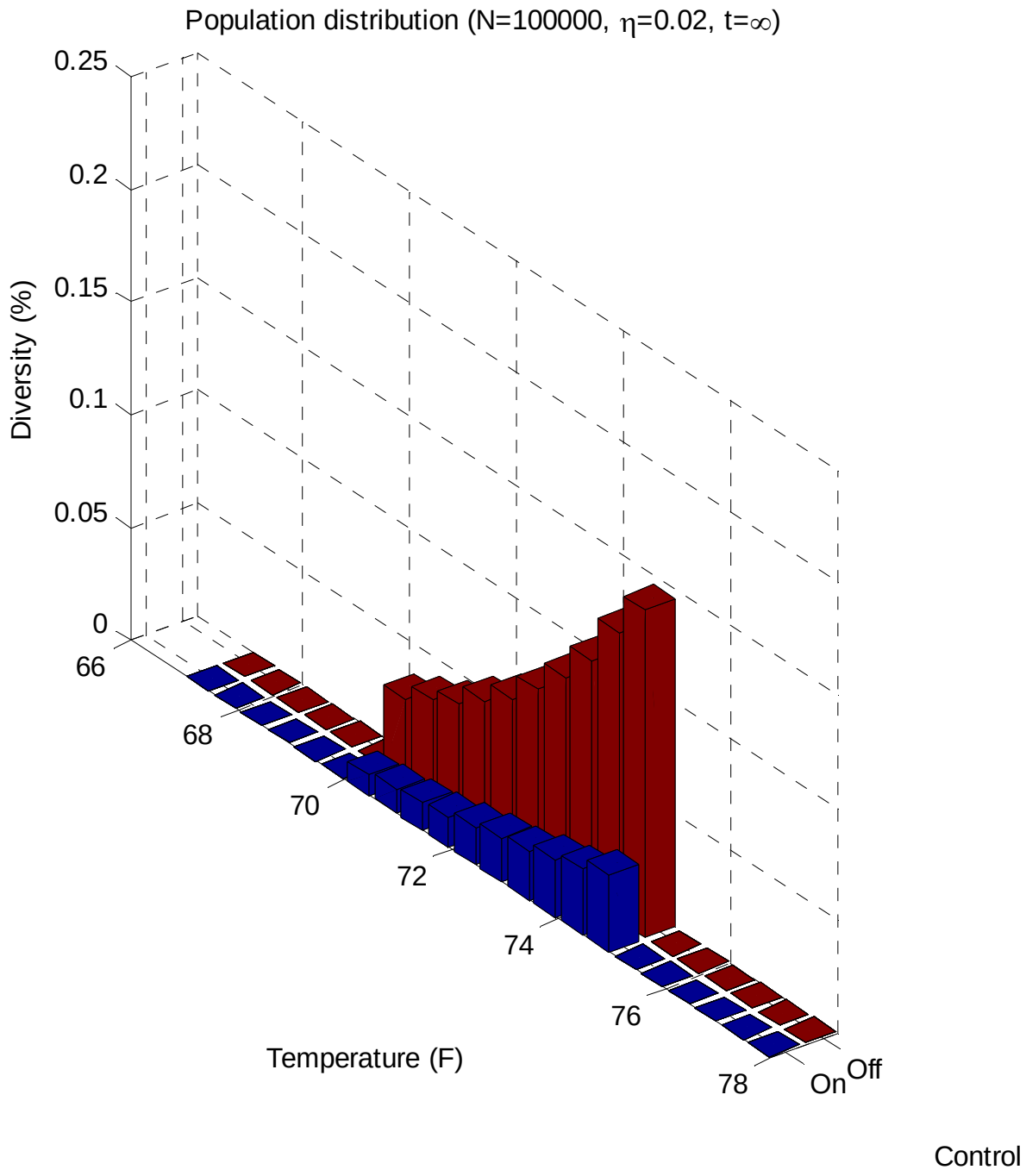

Fig. 3: The steady state distribution is not uniform when demand is non-zero

## IV. Analysis of State Dynamics

When a thermostatic end-use device is subjected to a demand, it seems we must consider the contribution that demand makes to the device's steady-state dynamics. In particular, it seems reasonable to consider that as the population of devices in the *off* mode cools down and approaches the *on* set-point $T_{on}$ they are uniformly being drawn from by demand. Those drawn upon by demand transition immediately to the *on* mode at a temperature located somewhere between the current temperature and $T_{on}$, depending on the type of end-use. The model studied assumes the temperature remains unchanged. The rate at which devices are drawn from the *off* mode is the demand rate, $\eta$. In the typical cycle of a thermostatic device with non-zero demand is illustrated. The number of devices *on* at a given temperature $T$ and a given instant t is denoted $n_{on}(T, t)$ with the second argument omitted in Figure 4 to emphasize that we are interested in steady-state equilibriums. Similarly for the number $n_{off}(T, t)$ of devices in the *off mode*.

The demand $\eta$ is given as an exogenous parameter that describes the rate at which appliances are subjected to demand events that would cause the controller to immediately transition to an *on* state. Note that $\eta$ affects all appliances regardless of whether they are *on* or *off* (because consumers do not choose to act in a way that creates demand on the basis of information regarding the current state of the appliance, nor do we desire that they should). The demand is a fractional rate between zero and one.

The functions $n_{off}(T,t)$ *and* $n_{on}(T,t)$ are related to one another and to the constants $r_{off}$, $r_{on}$, etc. by differential equations, which are themselves determined from difference equations that govern a quantized model. For clarity, let $n_x(T{:}dT,t)$ denote the number of devices in the mode $x \in \{on, off\}$ with temperatures between $T$ and $T+dT$ at the exact time $t$.

The variation $dn_x(T{:}dT)$ in the number of $x$-mode devices between time $t$ and $t+dt$ is caused by the combined effect of demand, of devices leaving the temperature range from $T$ to $T+dT$ due to natural cooling or heating, and of control changes of mode at set-point temperatures. The latter cause is captured in the boundary conditions. The critical quantities to estimate are the fraction of devices changing to the *on*-mode as a result of demand, the fraction of devices leaving the range $T$ to $T+dT$ during a give period of time $dt$ because of cooling or heating alone, and the impact of demand on the latter.

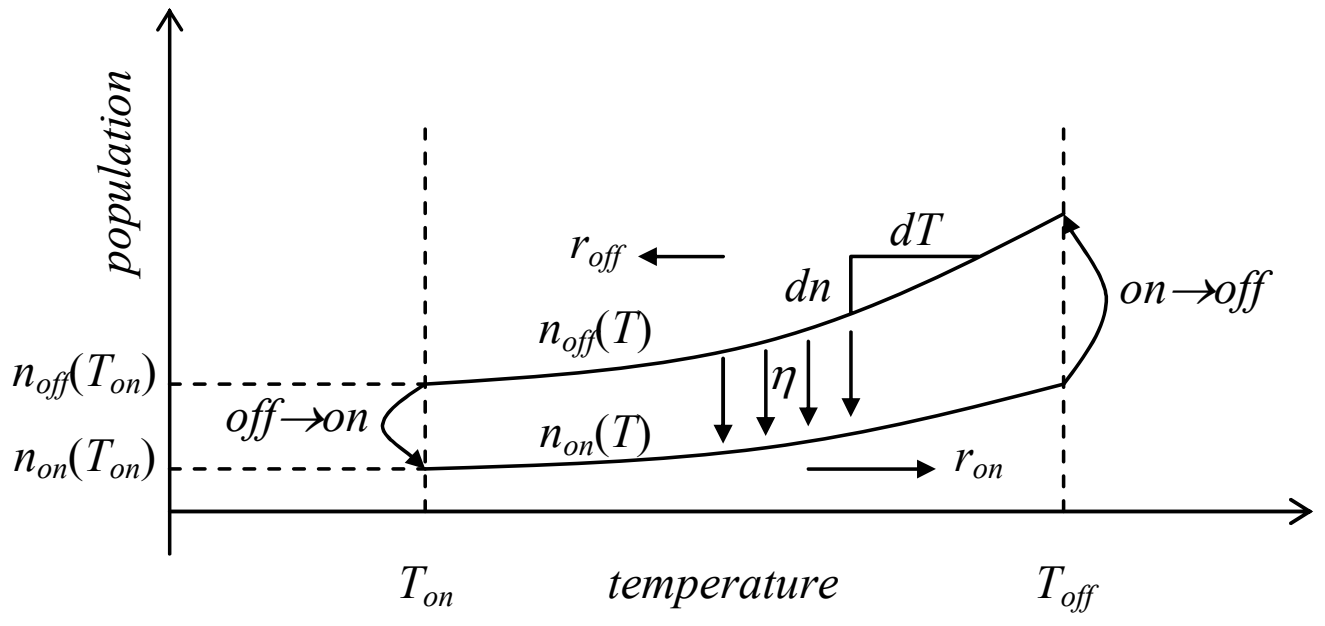

Fig. 4: The thermostatic end-use cycle for a heating regime with demand $\eta > 0$

Demand affects the *off* and *on* modes differently; it depopulates the *off*-mode states, and repopulates *on*-modes but affects only the instantaneous rate of heating. To make this statement precise, let $r_x$ denote the instantaneous rate at which the temperatures change in mode $x$ when demand is null; its units are in degrees per second (°C/sec) or some equivalent. The demand $\eta$ is the instantaneous probability of a random appliance usage event, independently of the initial mode of the device; its units may be given in $\sec^{-1}$.

$$\eta = \lim_{\Delta_t \to 0} \frac{prob(use \quad between \quad time \quad t \quad to \quad t+\Delta t)}{\Delta_t} \tag{6}$$

For now, let the given temperature $T$ be at least $2dT$ away from any temperature end-point. The distribution of temperatures is assumed to be approximately uniform within any temperature range of $dT$ units, that assumption is important but should be relative to easy to assert for large populations of machines as $dT \to 0$. During $dt$ time units and in the absence of demand, the devices in the *off*-mode see their temperature decrease by $dt \times r_{off}$ degrees. Thus at the end of the period $dt$, the range of temperatures for the devices accounted for in $n_{off}(T{:}dT,t)$ at time $t$ will extend to the left, as shown , by a proportion of $r_{off} \times dt/dT$. Under the assumption of a local uniform distribution, cooling alone removes approximately $(r_{off}\ dt/dT)\ n_{off}(T{:}dT,t)$ devices from the count $n_{off}(T{:}dT,t)$; and similarly for devices in the *on*-mode. The number of devices accounted for by $n_{off}(T{:}dT,t)$ that switch to *on*-mode due to demand in a duration of $dt$ is $\eta\ n_{off}(T{:}dT,t)\ dt$. All of this is summarized graphically in and in the finite difference equations (7).

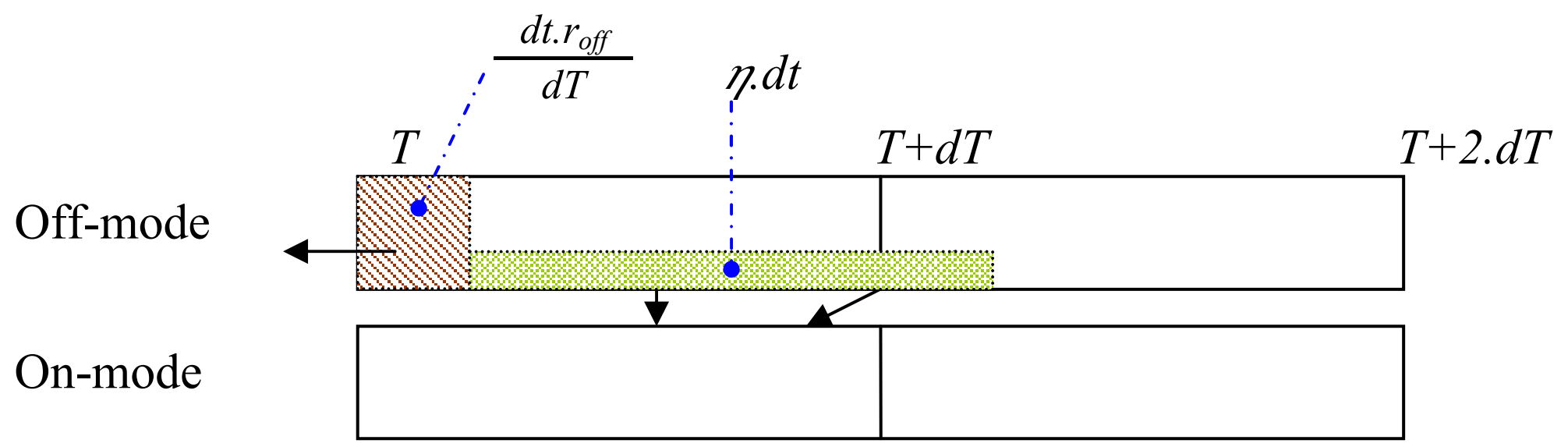

Fig. 5. The shaded areas represent devices that change their quantized sate between time t and t+dt.

$$\begin{aligned} dn_{off}(T:dT) \; = \; & -\eta.dt.n_{off}(T:dT,t) \\ & -\frac{r_{off}.dt}{dT}(1-\eta.dt).n_{off}(T:dT,t) \\ & +\frac{dt.r_{off}}{dT}(1-\eta.dt).n_{off}(T+dT:dT,t) \end{aligned} \qquad (7)$$

or equivalently

$$\begin{aligned} \frac{dn_{off}(T:dT)}{dt} \; = \; & -\eta.n_{off}(T:dT,t) \\ & +r_{off}\frac{n_{off}(T+dT:dT,t)-n_{off}(T:dT,t)}{dT}(1-\eta.dt) \end{aligned} \quad . \qquad (8)$$

A partial differential equation is obtained taking the limit of equation (8) as both $dt$ and $dT$ tend to zero, namely

$$\frac{\partial n_{off}}{\partial t} = -\eta.n_{off} + r_{off}\frac{\partial n_{off}}{\partial T} \qquad (9)$$

The equations for the *on*-mode devices, away from the temperature end points, are derived in a similar fashion; the main difference is in the effect of demand on the heating rate. Consider devices in the *on*-mode and whose temperature is within T and $T+\Delta T$. The average temperature of those devices is $T+½\Delta T$. At each instant during a period of time $dt$, there are about $\eta\ n_{off}(T{:}\Delta T,t)$ of those devices with a fixed temperature because of demand. The average temperature of those devices after $\Delta t$ time units is approximated by the sum of integrals

$$\eta\int_{s=T}^{T+\Delta t} s dT + (1-\eta)\int_{T}^{T+\Delta t}(s + r_{on}\Delta t)dT = T + \frac{\Delta T}{2} + (1-\eta)r_{on} \; . \qquad (10)$$

Effectively the instantaneous heating rate under demand is $r_{on}(1-\eta)$. The governing finite difference equations for on-mode devices away from temperature end points are therefore

$$\begin{aligned} dn_{on}(T:dT) \; = \; & \eta.dt.n_{off}(T:dT,t) \\ & -\frac{r_{off}(1-\eta).dt}{dT}.n_{on}(T:dT,t) \\ & +\frac{r_{off}(1-\eta)dt}{dT}.n_{on}(T-dT:dT,t) \end{aligned} \qquad (11)$$

The corresponding governing differential equation for *on*-mode devices is obtained by taking the limit of (11) as $dT$ and $dt$ approach zero, namely

$$\frac{\partial n_{on}}{\partial t} = \eta.n_{off} - r_{on}(1-\eta)\frac{\partial n_{on}}{\partial T} \; . \qquad (12)$$

## V. Time Independent Distributions

The time-independent solutions to this problem are obtained by setting to zero the time derivative given in (9) and (12), which leads us to a pair of ordinary differential equations in $T$:

$$\frac{\partial n_{off}}{\partial T} = \frac{\eta}{r_{off}} n_{off} \qquad (13)$$

$$\frac{\partial n_{on}}{\partial T} = \frac{\eta}{r_{on}(1-\eta)} n_{off} \qquad (14)$$

The time-independent functions $n_{off}$ and $n_{on}$ can be obtained, up to two constants $K_{off}$ and $K_{on}$, by integrating (13) and (14). Note that when $\eta=1$, Equation (12) forces $n_{off} = 0$.

$$n_{off} = K_{off} e^{\frac{\eta}{r_{off}}T} \qquad (15)$$

$$n_{on} = K_{off}\frac{r_{off}}{r_{on}(1-\eta)} e^{\frac{\eta}{r_{on}}T} + K_{on} \; . \qquad (16)$$

The boundary conditions for the time independent solutions in (15) and (16) simply states that inputs and outputs must match at the temperature end points, which takes the algebraic form

$$r_{on}(1-\eta)n_{on} = r_{off}n_{off} \tag{17}$$

It follows that $K_{on} = 0$. We also can compute $N_{off}$ and $N_{on}$, the total number of machines *off* and *on*, respectively, such that $N_{off} + N_{on} = N$, by integrating (16) and (17) between $T_{on}$ and $T_{off}$. The constant $K_{off}$ may be obtained from the equation for mass conservation (18).

$$\int_{T_{\min}}^{T_{\max}} n_{on} + n_{off} dT = \left(1 + \frac{r_{off}}{r_{on}(1-\eta)}\right)\int_{T_{\min}}^{T_{\max}} n_{off} dT = N \ . \tag{18}$$

Explicitly, one finds

$$K_{off} = \frac{N\eta(1-\eta)r_{on}}{\left[(1-\eta)r_{on} + r_{off}\right]\left[e^{\frac{\eta}{r_{off}}T_{\max}} - e^{\frac{\eta}{r_{off}}T_{\min}}\right]} \ . \tag{19}$$

Finally, we can compute the load diversity by integrating (17) to derive a linear relationship between $N_{off}$ and $N_{on}$ and substituting into the definition of $\varphi$.

$$\begin{aligned} \varphi &= \frac{N_{on}}{N_{off} + N_{on}} \\ &= \frac{r_{off} / r_{on}(1-\eta)}{1 + r_{off} / r_{on}(1-\eta)} \\ &= \frac{r_{off}}{r_{off} + (1-\eta)r_{on}} \end{aligned} \tag{6}$$

This result is compared to the diversity observed in the simulation, as shown in Fig. 6. The deviation of the model from the simulation is one of the principal flaws of numerical simulations of this type. The discrete nature of the simulation can result in variations in the results for certain parameters.

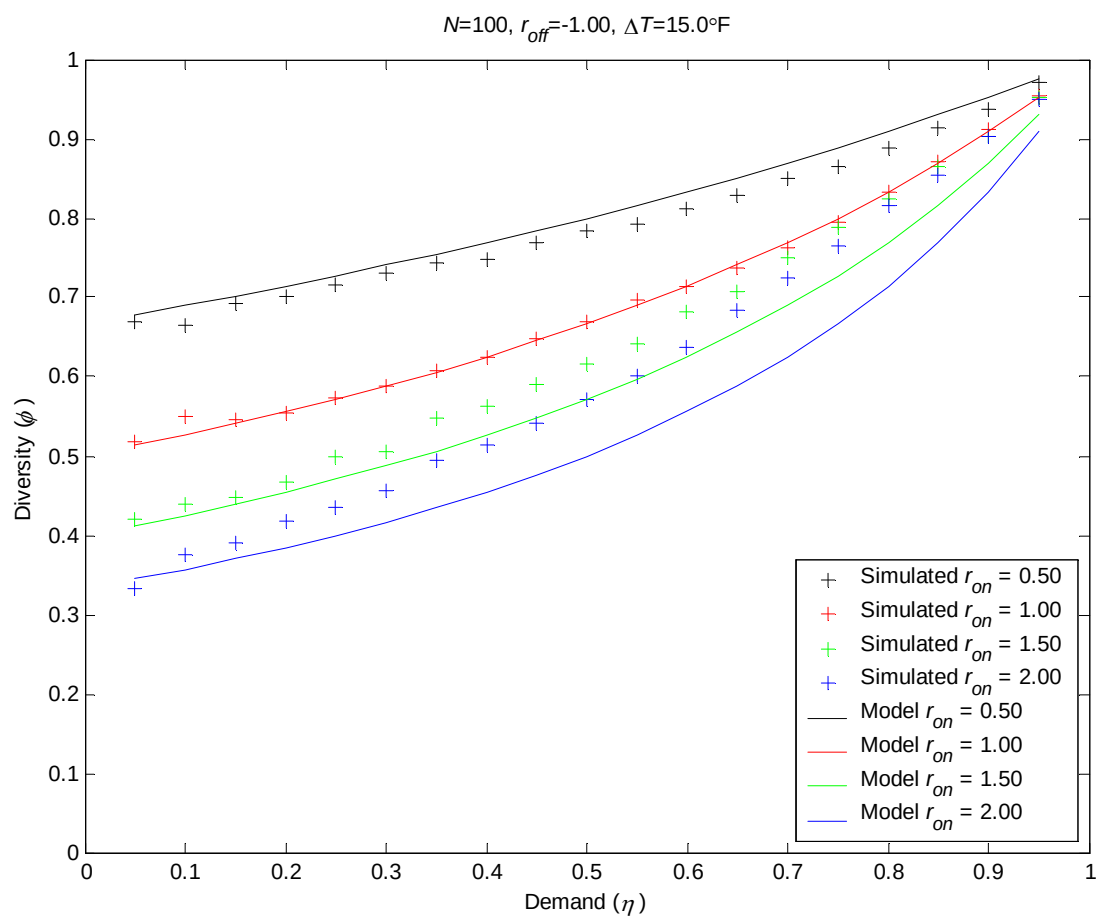

Fig. 6: The expected load diversity is compared to a simulation of 100 waterheaters.

It would seem that superficially this result does not greatly affect load forecasting and distribution system planning to any great extent. As previously noted, the impact of incorrectly computing the steady-state distribution resulted only in an initialization transient in the simulation (a rather common occurrence) and a small error in the expected diversity. However, it has already been shown that price fluctuations can dramatically impact load diversity and that demand plays a role in dampening that phenomenon [10]. Furthermore, we have treated demand as a time-independent constant. The natural cycling period of the devices under consideration is often on the order of the period of diurnal demand cycles and price fluctuations. Thus the above consideration would benefit greatly from the inclusion of a more general time-dependent demand function similar to those observed in end-use metered data [15]. It is reasonable to believe that such a refinement should lead to much better insights into the relationship between price fluctuations and load diversity.

## VI. Time Dependent Distributions

First, the only distributions that satisfy the flow equations (17) at all temperatures are time-independent. To see this, we let $q$ and $Q$ be the two variables defined implicitly by the relations

$$\begin{array}{rcl} t &=& -q + Q \\ T &=& r_{off}.q + r_{on}(1-\eta)Q \end{array} \ . \tag{21}$$

We can write (9) and (12) in terms of $q$ and $Q$ as Equations (22) and (23).

$$\frac{\partial n_{off}}{\partial q} = -\frac{\partial n_{off}}{\partial t} + r_{off}\frac{\partial n_{off}}{\partial T} = \eta . n_{off} \quad (22)$$

$$\frac{\partial n_{on}}{\partial Q} = \frac{\partial n_{on}}{\partial t} + r_{on}(1-\eta)\frac{\partial n_{on}}{\partial T} = \eta . n_{off} \quad (23)$$

Equation (22) gives immediately an expression for $n_{off}$, namely

$$n_{off} = k_{off}(Q).e^{\eta \cdot q}, \quad (24)$$

where $k_{off}(Q)$ is an unknown function of $Q$ alone. Combining the boundary conditions (17), with (23) and (24) gives

$$\frac{\partial n_{on}}{\partial Q} = \frac{r_{off}}{r_{on}(1-\eta)}.e^{\eta q}\frac{\partial k_{off}(Q)}{\partial Q} = \eta . k_{off}(Q)e^{\eta q} \quad (26)$$

Considering the second equality, leads to the differential equation

$$\frac{\partial k_{off}(Q)}{\partial Q} = \eta \frac{r_{on}(1-\eta)}{r_{off}} k_{off}(Q), \quad (27)$$

whose solution up to a constant $h$ is

$$k_{off}(Q) = h.e^{\eta(1-\eta)\frac{r_{on}}{r_{off}}Q}. \quad (28)$$

and can be substituted back into (24) to give

$$n_{off}(q,Q) = h.e^{\frac{\eta}{r_{off}}\left(Q(1-\eta).r_{on} + q.r_{off}\right)}, \quad (29)$$

and after substitution of t and $T$ for $q$ and $Q$

$$n_{off}(t,T) = h.e^{\frac{\eta}{r_{off}}T}. \quad (30)$$

The latter is independent of time and so must be $n_{on}$.

## VII. Conclusion

We have shown that the standard model of load diversity fails to accurately predict the steady-state diversity observed in highly accurate simulations of end-use load behavior. The discrepancies, when examined closely suggest that the steady-state distribution of end-use device states is not uniform as generally assumed. Consideration of this fact has led to a more accurate expression of the steady-state distribution of end-use device states. This can be used to correctly initialize agent-based end-use simulations and accurately compute the expected load diversity for any given level of end-use demand. It is also expected to provide a better understanding of the steady-state regime of end-use loads and ultimately lead to an improved generalized model of the time-dependent behavior of loads responding to changes in demand and set-points such as might be implemented in real-time price control strategies.

## VIII. Acknowledgements

The author would like to thank Jeffry V. Mallow of Loyola University Chicago for his generous and insightful contributions. This work was funded by U.S. Department of Energy's Pacific Northwest National Laboratory under the Energy Systems Transformation Initiative. Pacific Northwest National Laboratory is operated by Battelle Memorial Institute under Contract DE-ACO6-76RL01830.

## IX. References

[1] L.D. Kannberg, D.P. Chassin, J.G. Desteese, S.G. Hauser, M.C. Kintner-Meyer, R.G. Pratt, L.A. Schienbein, and W.M. Warwick, *GridWise™: The Benefits of a Transformed Energy System*, Report no. PNNL-14396, Pacific Northwest National Laboratory, Richland, Washington, September 2003.

[2] J. Eto, C. Marnay, C. Goldman, J. Kueck, J. Dagle, F. Alvarado, T. Mount, S. Oren, and C. Martinez, "An R&D Agenda to Enhance Electricity System Reliability by Increasing Customer Participation in Emerging Competitive Markets", *Power Engineering Society Winter Meeting*, vol. 1, pp.247-251, 2001.

[3] R.E. Bohn, *Spot Pricing of Public Utility Service*, PhD. dissertation, MIT Energy Laboratory Technical Report MIT-EL-82-031, 1982.

[4] F.C. Schweppe, B. Darayian, and R.D. Tabors, "Algorithms for a Spot Price Responding Residential Controller", *IEEE Transactions on Power Systems*, vol. 4, pp. 507-516, 1989.

[5] B. Darayian, R. E. Bohn, and R. D. Tabors, "Optimal Demand-side Response to Electricy Spot Prices for Storage-type customer", *IEEE Transactions on Power Systems*, vol. 4, pp. 897-903, 1989.

[6] D.P. Chassin, "Statistical mechanics: a possible model for market-based electric power control", in proc. of *37th Hawaii International Conference on Systems Science* (HICSS-37), 2004.

[7] R.T. Guttromson, D. P. Chassin, and S. E. Widergren, "Residential Energy Resource Models for Distribution Feeder Simulation", *Proc. of 2003 IEEE PES General Meeting*, Toronto, Canada, 2003.

[8] Z.T. Taylor, R.G. Pratt, "Description of Electric Energy End-use in Commercial Buildings in the Pacific Northwest", report no. DOE/BP-13795-22: *End-use Load and Consumer Assessment Program (ELCAP)*, Bonneville Power Administration, 1989.

[9] Z.T. Taylor, and R.G. Pratt, "The Effects of Model Simplification on Equivalent Thermal Parameters Calculated from Hourly Building Performance Data", in *American Council for an Energy Efficient Economy (ACEEE) 1988 Summer Study on Energy Efficiency in Buildings*, vol. 10, proc. from the Panel on Performance Measurement and Analysis, 1988.

[10] N. Lu, Z. T. Taylor, D. P. Chassin, R. T. Guttromson, and R. S. Studham, "Parallel computing environments and methods for power distribution system simulation", submitted to HICSS-38, 2004.

[11] N. Lu, and D. P. Chassin, "A State Queueing Model of Price Responsive Cyclic Loads", *IEEE Transactions on Power Systems*, publication due in 2004.

[12] N. Ryan, S.D. Braithwait, J.T. Powers, and B.A. Smith, "Generalizing Direct Load Control Program Analysis: Implementation of the Duty Cycle Approach", *IEEE Transactions on Power Systems*, vol. 4, no. 1, pp. 293-299, 1989.

[13] A.K. Glosh, D.L. Lubkeman, M.J. Downey, and R.H. Jones, "Distribution Circuit State Estimation Using a Probabilistic Approach", *IEEE Transaction on Power Systems*, vol. 12, no. 1, pp. 45-51, 1997.

[14] N.D. Hatziargyriou, T.S. Karakatsanis, and M. Papadopoulos, "Probabilistic Calculations of Aggregate Storage Heating Loads", *IEEE Transactions on Power Delivery*, vol. 5, no. 3, pp. 1520-1526, 1990.

[15] R.G. Pratt, C.C. Conner, E.E. Richman, K.G. Ritland, W.F. Sandusky, and M.E. Taylor, "Description of Electric Energy Use in Single-family Residences in the Pacific Northwest," *End-use Load and Consumer Assessment Program (ELCAP)*, report to Bonneville Power Administration, Pacific Northwest National Laboratory, Richland, Washington, July 1989.

## X. Biographies

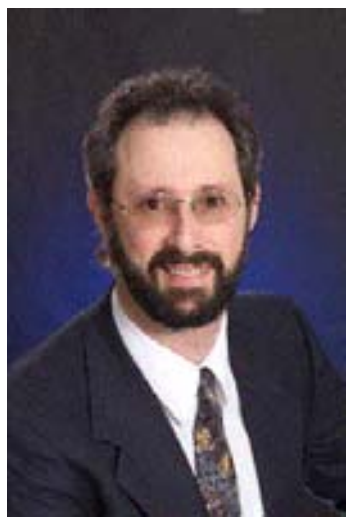

**David P. Chassin** (M'2003) received his BS of Building Science from Rensselaer Polytechnic Institute in Troy, New York. He is staff scientist with the Energy Science & Technology Division at Pacific Northwest National Laboratory where he has worked since 1992. He was Vice-President of Development for Image Systems Technology from 1987 to 1992, where he pioneered a hybrid raster/vector computer aided design (CAD) technology called CAD Overlay™. He has developed building energy simulation and diagnostic systems, leading the development of Softdesk Energy and DOE's Whole Building Diagnostician. He has served on the International Alliance for Interoperability's Technical Advisory Group, chaired the Codes and Standards Group. His recent research focuses on emerging theories of complexity and their applications to high-performance simulation and modeling of engineered systems.

**Joel Malard** received his PhD from the School of Computer Science at McGill University in 1993, having developed some very efficient QR factorization algorithms for medium size matrices using collective communications on multicomputers. He is a Scientist IV within the Computational Sciences & Mathematics Division Pacific Northwest National Laboratory where he has worked since 1997. There he has worked on parallel iterative solvers for linear systems arising in subsurface flow simulations, global optimization using evolutionary computing as well as developing numerical algorithms for statistical computations in bio-informatics and micro-array data analysis. Joel has been application scientist at the Edinburgh Parallel Computing Centre in Scotland from 1994 to 1997, his MSc is in Algebraic Geometry from McGill University. Joel is particularly interested in integer and floating point computations that outstretch the capabilities of current high-performance computing platforms.

Filename: Chassin and Malard, Equilibrium dynamics of thermostatic end-use load diversity (Final).doc
Directory: \\simon\my documents\ESTI\FY04 LDRD\Load Diversity
Template: C:\Documents and Settings\d3g637\Application Data\Microsoft\Templates\Normal.dot
Title: The Equilibrium Dynamics of Thermostatic End-Use Load Diversity as a Function of Demand
Subject:
Author: David P. Chassin
Keywords:
Comments:
Creation Date: 7/29/2004 1:20 PM
Change Number: 13
Last Saved On: 9/3/2004 4:01 PM
Last Saved By: David P. Chassin
Total Editing Time: 176 Minutes
Last Printed On: 9/18/2004 9:11 AM
As of Last Complete Printing
Number of Pages: 8
Number of Words: 5,651 (approx.)
Number of Characters: 32,217 (approx.)